\documentclass[twocolumn,numberedappendix]{openjournal}
\usepackage{anyfontsize}
\usepackage{graphicx} % Required for inserting images
\usepackage{tikz,xspace}
\usetikzlibrary{patterns, snakes}
\usetikzlibrary{positioning}

\begin{document}
\title{WHFast512: A symplectic N-body integrator for planetary systems optimized with AVX512 instructions}
\shorttitle{WHFast512 - N-body integrator optimized with AVX512 instruction}
\shortauthors{Javaheri, Rein \& Tamayo}

\author{Pejvak Javaheri}
\affiliation{Department of Physical and Environmental Sciences, University of Toronto at Scarborough, Toronto, Ontario M1C 1A4, Canada}

\author{Hanno Rein}
\affiliation{Department of Physical and Environmental Sciences, University of Toronto at Scarborough, Toronto, Ontario M1C 1A4, Canada}
\affiliation{Department of Astronomy and Astrophysics, University of Toronto, Toronto, Ontario, M5S 3H4, Canada}

\author{Daniel Tamayo}
\affiliation{Department of Physics, Harvey Mudd College, 301 Platt Blvd., Claremont 91711, USA\\}

\begin{abstract}
    We describe the implementation of the symplectic N-body integrator WHFast512 using Single Instruction Multiple Data (SIMD) parallelism and 512-bit Advanced Vector Extensions (AVX512). 
    We are able to speed up integrations of planetary systems by up to 4.7x compared to the non-vectorized version of WHFast. 
    WHFast512 can integrate the Solar System with 8 planets for 5~billion years in less than 1.4~days.
    To our knowledge, this makes WHFast512 the fastest direct N-body integrator for systems of this kind. 
    As an example, we present an ensemble of 40-Gyr integrations of the Solar System.
    Ignoring the Sun's post-main sequence evolution, we show that the instability rate is well captured by a diffusion model.
    WHFast512 is freely available within the REBOUND package.
\end{abstract}

\maketitle

\section{Introduction}
\label{sec:introduction}
Numerical integrations of the Solar System have a long history \citep[for a historical summary, see e.g.][]{Laskar2013}.
With the discovery of extrasolar planetary systems, the interest in fast and efficient numerical methods to study orbital evolution has further increased.

Most direct N-body simulations use symplectic integrators that make use of the so-called Wisdom-Holman splitting \citep{Wisdom1981, WisdomHolman1991, Kinoshita1991}.
One frequently used implementation of a Wisdom-Holman integrator is WHFast \citep{ReinTamayo2015}, which is freely available within the REBOUND package \citep{ReinLiu2012}.
As the name suggests, WHFast was developed with efficiency being one of the goals. 

WHFast and most other N-body integrators used for long-term integrations of planetary systems run on CPUs.
Other astrophysical simulations have been very successful in achieving significant speed-ups using Graphics Processing Units (GPUs).  
This not only includes cosmology or fluid dynamics codes but also N-body integrators \citep[see e.g][]{Grimm2022}.
The reason why long-term integrations of planetary systems are typically not run on GPUs is that they are inherently sequential: timesteps are performed one after the other and during each timestep each body depends on the positions of all the other bodies.
In simulations with a small number of particles (e.g., 1 star and 8 planets), the time required to perform each timestep is short, typically of the order of microseconds. 
Almost any attempt to make use of the parallel computing power of a GPU to accelerate a single timestep is going to experience a slowdown rather than a speedup because of the communication overhead that occurs even on shared memory systems\footnote{See \cite{SahaStadelTremaine1997} for an alternative approach which does provide a significant speedup of 50x but increases the computational cost by a factor of 1000.}.
CPUs typically have higher clock rates than GPUs and can therefore finish simulation faster.

Because one needs to resolve at least the orbital timescale of the short-period planet with several timesteps, the total number of timesteps in a long-term integration can be \emph{very} large, $\gtrsim 10^{12}$.
As a result, a single integration might run for weeks or months.
Of course, GPU or CPU-based computing clusters allow for many independent integrations to be performed in parallel.
But this leaves us in an unusual situation where someone might have access to a vast amount of computing power, but without a way to use those resources to accelerate a single integration.

In this paper, we describe a way to speed up small N-body simulations with 512-bit wide Advanced Vector Extensions (AVX512). 
Our new integrator can speed up both a single integration and an ensemble of integrations by almost a factor of 5.

We first provide a short introduction to AVX512 in Section~\ref{sec:avx512}.
We then give a review of Wisdom-Holman integrators and the specific Hamiltonian splitting we're using, before providing a detailed description of the implementation of WHFast512 in Section~\ref{sec:whfast512}.
Performance tests are described in Section~\ref{sec:performance}.
As an example, we present the results of an ensemble of 40~Gyr integrations of the Solar System in Section~\ref{sec:results}.
We conclude in Section~\ref{sec:conclusions} with some ideas for further optimizations.

\section{Advanced Vector Extensions}
\label{sec:avx512}
AVX512 instructions are an extension to the x86 CPU instruction set.
They are available on some, mostly high-end, Intel and AMD CPUs. 
It allows programs to pack eight 64-bit double-precision floating point numbers within 512-bit vectors and perform operations on these vectors.
One AVX512 instruction can perform as many double floating point operations as eight regular instructions.
In an ideal scenario we might achieve a speed-up of~8x.

Consider the task of adding an eight element vector $(a_0,\ldots, a_7)$ to another vector $(b_0,\ldots , b_7)$. This can be done with the AVX512 instruction \texttt{vaddpd}:
\begin{center}
\begin{tikzpicture}[scale=0.65]
\tikzstyle{every node}=[font=\fontsize{8}{8}\selectfont]

\begin{scope}[shift={(0,0)}]
    \draw[xstep=15mm,black] (0,0) grid (8*1.5,1);
    \foreach \i in {0,...,7}{
        \node at (0.75+1.5*\i,0.5) {$a_\i$};
    }
\end{scope}
\foreach \x in {0,...,7}{ 
    \draw[color=black,fill=white] (1.5*\x+0.75,-0.5) node {+};
}
\begin{scope}[shift={(0,-2)}]
    \draw[xstep=15mm,black] (0,0) grid (8*1.5,1);
    \foreach \i in {0,...,7}{
        \node at (0.75+1.5*\i,0.5) {$b_\i$};
    }
\end{scope}
\foreach \x in {0,...,7}{ 
    \draw[color=black,fill=white] (1.5*\x+0.75,-2.5) node {=};
}
\begin{scope}[shift={(0,-4)}]
    \draw[xstep=15mm,black] (0,0) grid (8*1.5,1);
    \foreach \i in {0,...,7}{
        \node at (0.75+1.5*\i,0.5) {$a_\i{+}b_\i$};
    }
\end{scope}
\end{tikzpicture}
\end{center}
The example above operates on the two input vectors vertically.
There are some instructions such as permutations that can operate on vectors horizontally. 
For example, the AVX512 instruction \texttt{vshufpd} can be used to shuffle neighbouring elements in a vector $(a_0,\ldots,a_7$):
\begin{center}
\begin{tikzpicture}[scale=0.65]
\tikzstyle{every node}=[font=\fontsize{8}{8}\selectfont]

\begin{scope}[shift={(0,0)}]
    \draw[xstep=15mm,black] (0,0) grid (8*1.5,1);
    \foreach \i in {0,...,7}{
        \node at (0.75+1.5*\i,0.5) {$a_\i$};
    }
\end{scope}
   \foreach \x in {0,...,3}{   
        \draw[<-] (1.5*2*\x+0.75,-0.8) -- (1.5*2*\x+2.25,-0.2);
        \draw[<-] (1.5*2*\x+2.25,-0.8) -- (1.5*2*\x+0.75,-0.2);
    }
\begin{scope}[shift={(0,-2)}]
    \draw[xstep=15mm,black] (0,0) grid (8*1.5,1);
    \foreach \i in {0,...,7}{
        \pgfmathparse{Mod(\i,2)==0?1:0}
        \ifnum\pgfmathresult>0
            \node at (0.75+1.5*\i+1.5,0.5) {$a_\i$};
        \else
            \node at (0.75+1.5*\i-1.5,0.5) {$a_\i$};
        \fi
    }
    \foreach \i in {0,...,3}{
        \draw [thick, decoration={brace, mirror, raise=0.5cm},decorate] (0.1+2*\i*1.5,0.5) -- (1.9*1.5+2*\i*1.5,0.5); 
        \node at (1*1.5+2*\i*1.5,-0.7) {128-bit lane};
    }
    \foreach \i in {0,...,1}{
        \draw [thick, decoration={brace, mirror, raise=0.5cm},decorate] (0.1+4*\i*1.5,-0.5) -- (3.9*1.5+4*\i*1.5,-0.5); 
        \node at (2*1.5+4*\i*1.5,-1.7) {256-bit lane};
    }
\end{scope}
\end{tikzpicture}
\end{center}
Depending on the specific CPU architecture, some horizontal operations involve a performance penalty when \textit{lanes} are crossed, especially when crossing from one 256-bit lane to another.  
In the above example using the \texttt{vshufpd} instruction neither 128-bit nor 256-bit lane boundaries are crossed.
This is important because in an N-body simulation every planet interacts with every other planet, so at some point we need to transfer information across lanes. 
As we describe in more detail below, we try to minimize these lane crossings for the Kepler and jump steps in WHFast512.

We are specifically interested in creating a fast integrator for systems with 8 planets.
With this in mind, it makes sense to distribute particle data in 512-bit vectors, each containing one element of each planet.
We have one vector for the masses of each planet, one for the x-coordinate of each planet, one for the y-coordinates, and so on. 

\section{WHFast512}
\label{sec:whfast512}
In this section, we introduce the WHFast512 algorithm. 
We provide a short general introduction to Wisdom Holman integrators before discussing specifically WHFast512 and its implementation.

\subsection{Wisdom-Holman integrator}
The main idea behind the Wisdom-Holman (WH) integrator is to split the complicated orbital motion of planets into smaller, more tractable parts \citep{Wisdom1981,WisdomHolman1991,Kinoshita1991}. 
As long as planets remain well separated, the gravitational interactions between a planet and the star dominate the motion and planet-planet interactions can be considered a perturbation.
We can stitch together an approximation to the full solution by frequently switching back and forth between propagating planets on Keplerian ellipses around the star (ignoring planet-planet interactions) and applying kicks due to other planets (ignoring the motion around the star). 

Wisdom-Holman integrators can be used with a variety of different coordinate systems.
The choice of the coordinate system affects the choice of Hamiltonian splitting. 
For example, the WHFast implementation in REBOUND supports Jacobi coordinates, democratic heliocentric coordinates \citep{Duncan1998}, and the WHDS splitting \citep{HernandezDehnen2016}.
There are advantages and disadvantages to each splitting. 
See \cite{ReinTamayo2019} for an overview.
For reasons we will explain in more detail below, we here use the Wisdom-Holman integrator with democratic heliocentric coordinates (WHD).
In these coordinates, the position of particle $i$ with mass $m_i$ is $\mathbf Q_i$ and is measured relative to the star.
The corresponding canonical momentum $\mathbf P_i$ is measured relative to the barycentre. 
$\mathbf P_0$ and $M$ are the total momentum and total mass of the system respectively. 

Following \cite{Duncan1998}, we write the N-body Hamiltonian in democratic heliocentric coordinates and split it into four parts:
\begin{eqnarray}
    H &=& \underbrace{
     \vphantom{ \sum_{i=1}^N} 
    \frac{P_0^2}{2M}}_{H_0}
    \;
    \underbrace{+\sum_{i=1}^N \left( \frac{P_i^2}{2m_i} - \frac{Gm_0m_i}{Q_i} \right)}_{H_K}
    \;\nonumber
    \\
    &&
    \underbrace{
    \vphantom{\left( \sum_{i=1}^N  \mathbf P_i \right)^2 } 
    -\sum_{i=1}^N\sum_{j=i+1}^N \frac{Gm_i m_j}{Q_{ij}}}_{H_I}
    \;
    \underbrace{
    \vphantom{\sum_{j=i+1}^N \frac{Gm_i m_j}{Q_{ij}}} 
    + \frac1{2m_0} \left( \sum_{i=1}^N  \mathbf P_i \right)^2}_{H_J}.
\end{eqnarray}
where $\mathbf Q_{ij} = \mathbf Q_i - \mathbf Q_j$.
$H_0$ describes the motion of the barycenter. 
Because there are no external forces, the barycenter travels along a straight line with constant velocity.
This part is therefore trivial to solve.
$H_K$ describes the Keplerian motion of each planet. 
$H_I$ describes the planet-planet interactions.
Because we are using democratic heliocentric coordinates, we have an additional term, $H_J$, the so-called jump step. 
During this step, the positions of all planets change by the same amount, but the momenta stay constant (hence the name \textit{jump step}).

One full Wisdom-Holman timestep of length $dt$ in operator notation is then given by
\begin{equation}
   \hat H_K\left(\frac{dt}2\right) 
     \cdot \hat H_J\left(\frac{dt}2\vphantom{\frac{dt}2 }\right)  
      \cdot \hat H_I\left(dt\vphantom{\frac{dt}2 }\right) 
      \cdot \hat H_J\left(\frac{dt}2\vphantom{\frac{dt}2 }\right)  
      \cdot \hat H_K\left(\frac{dt}2\right)
\end{equation}
where we are ignoring $\hat H_0$ as it commutes with all other operators. 
Unless we require an output, the first and last Kepler steps (which are the most computationally expensive) can be combined.
Furthermore, the jump step commutes with the interaction step if no general relativistic corrections are included (see below). If that is the case, then the two jump steps can be combined.
The above splitting scheme results in a second-order integrator where the energy error scales as $O(\epsilon\, dt^2)$ where $\epsilon$ is of the order of~$m_i/M$.
Higher order generalizations of the WH integrator exist but are not considered in this paper \citep[see][for an overview]{ReinTamayo2019}.
The individual steps are described in detail in the following sections. 

\subsection{Kepler Step}
\label{sec:kepler}

\begin{figure*}
    \centering
    \includegraphics[width=\textwidth]{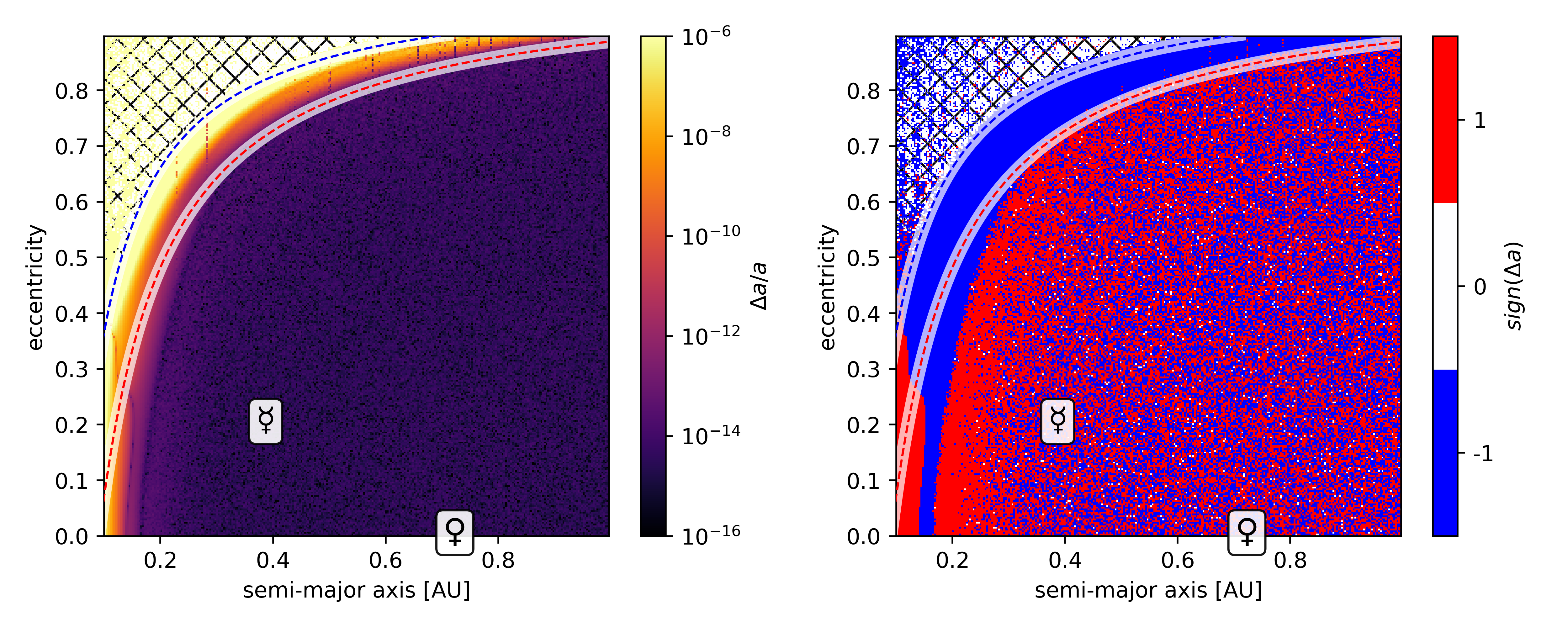}
    \caption{This plot shows the relative error of the semi-major axis for a test particle evolved for 10 years as a function of the semi-major axis and eccentricity. The left-hand side shows the absolute error in the semi-major axis. The right-hand side shows the sign of the error.
    The present-day orbital elements of Mercury and Venus are shown for comparison.
    On the dashed blue line, the pericenter timescale is equal to the timestep, on the red dashed line, it is equal to 2 timesteps.
    The hatched style on the top left indicates where NaN~values (not a number) occur.
    }
    \label{fig:kepler}
\end{figure*}

During the Kepler step, particles are moved along their Keplerian orbits. 
These Keplerian orbits are independent of each other because no planet-planet interactions occur during this step.
It is therefore ideal for parallelizing under a single instruction multiple data (SIMD) paradigm.
Our new Kepler step is based on WHFast but with several modifications and limitations. 
The main reason for these modifications is that we want to avoid any conditional branching during the step to keep the calculations of all eight planets in sync.

For each planet, we need to solve the equations of motion corresponding to the Hamiltonian 
\begin{equation}
    H_{K,i} = \frac{P_i^2}{2m_i} - \frac{Gm_0m_i}{Q_i}.
\end{equation}
To obtain a solution we need to first solve Kepler's equation.
We do this iteratively using a combination of Halley's method and Newton's method. 
In the original version of WHFast, the iteration is stopped when the solution has converged to machine precision or begins to oscillate. 
Here, we use a fixed number of two Halley steps followed by two Newton steps. 
For low eccentricity elliptical orbits and small timesteps, this is sufficient to achieve machine precision. 
However, the solution might not converge for highly eccentric or hyperbolic orbits. 
Furthermore, when the timestep becomes comparable to the orbital period (or more precisely the pericentre timescale, see below), then the solution might also not converge.

Instead of cosine and sine functions, the Kepler solver in WHFast uses Stumpff and Stiefel functions. 
These are evaluated using a Taylor series expansion and a precomputed lookup table. 
In the standard version of WHFast, it is guaranteed that the series converges by repeatedly applying the quarter angle formula.
For WHFast512, we do not guarantee this because it would require branching. Instead, we assume that the timestep is always small compared to the orbital period \citep[specifically, we assume that $X\lesssim 0.1$, see][for the definition of $X$]{ReinTamayo2015}.

In the original WHFast version, the number of terms in the Taylor series expansion of the Stumpff function is fixed to 6.
Here, we use only 4 terms during the first two steps of the iteration (Halley's method).
During the last two steps in the iteration (Newton's method), we use 8 terms. 
Our tests (see below) show that a less accurate result for the first two iterations are good enough to improve the initial guess. 
The last iterations can then further increase the accuracy and reach machine precision.
For details on how we arrived at this specific combination, see Appendix~\ref{sec:iterations}.

We now show that all modifications mentioned above do not affect the accuracy of the Kepler solver as long as we know that the orbital periods in the planetary system don't become too short and that the orbits don't become too eccentric during the integration for a given timestep.

To test where our Kepler step fails, we run integrations of a massless test particle with varying semi-major axis and eccentricity. 
The semi-major axis should be conserved exactly in this test problem\footnote{A two-body problem where both particles have a finite mass does not conserve the semi-major axis exactly because of our choice of Hamiltonian splitting which includes a jump term.}.
We plot the magnitude and sign of the relative error in semi-major axis after ten years in Fig.~\ref{fig:kepler}.
A 5-day timestep is used in these integrations and the initial phase of the test particle is random. 
On the dashed blue line, the pericenter timescale, $T_f$, defined as
\begin{equation}
T_f = \frac{2\pi}n \frac{(1-e)^2}{\sqrt{1-e^2}},    \label{eq:peri}
\end{equation}
where $n$ and $e$ are the planet's mean motion and eccentricity, respectively, 
is equal to the timestep. 
On the red dashed line, $T_f$ is equal to two timesteps. 
Note that $T_f$ is simply the orbital period for circular orbits. For eccentric orbits it provides a timescale that describes how long a particle spends near pericenter. \cite{Wisdom2015} calls this the \textit{effective} period at pericenter.

One can see that with a 5-day tiemstep, our Kepler solver achieves machine precision accuracy for test particles with Mercury's semi-major axis up to eccentricities of $\sim 0.7$. 
Furthermore, note that the sign of the error is random whenever the result is converged, indicating that the solver is unbiased.
We therefore don't expect a drift in the semi-major axis over long timescales as long as Mercury's semi-major axis doesn't shrink significantly and we always resolve $T_f$ with at least two timesteps.
Note that this discussion so far only applies to the convergence of the Kepler solver - a smaller timestep might be needed to accurately integrate the system (see the discussion in Section~\ref{sec:results}).

\subsection{Interaction Step}
In the interaction step, we solve the equations of motion corresponding to the Hamiltonian $H_I$.
For each planet, the solution involves calculating the sum of the accelerations from all other planets, specifically
\begin{equation}
 \mathbf a_i = -\sum_{j=1,\, j\neq i}^N \frac{G m_j}{Q^3_{ij}} \mathbf Q_{ij} .\label{eq:acc}
\end{equation}
Once the accelerations are calculated, we only need to multiply them with the timestep and then add the result to the velocities.
Nevertheless, the interaction step is more complicated to implement because for every particle we need information from all other particles to calculate the acceleration. 
Thus, this step does not follow a simple SIMD paradigm like the Kepler step.

We restrict our discussion to the case of one~star and eight~planets. 
There are a total of $8\cdot(8-1)$ acceleration terms to calculate. 
Because of Newton's third law of motion, we can reduce the number of square root and inverse operations by half to $8\cdot(8-1)/2=28$.
In addition to the square root and inverse operations, we also need to perform other operations, for example to calculate $\mathbf{Q}_{ij}$.
However, the square root and inverse calculations are by far the most time consuming operations\footnote{An add, multiply, or fused multiply-add operation takes 4 clock cycles whereas a square root or inverse operation takes 31 or 23 clock cycles respectively on an Intel Skylake CPU.}.
In the standard version of WHFast, the 28 square roots and inverse operations are calculated sequentially. 
Using AVX512, we can calculate up to eight double-precision square roots or inverse calculations with one instruction.
For eight planets, we thus only need four square-root instructions to calculate all accelerations. 

We now discuss the specific order in which we calculate the pairwise accelerations. 
Although different choices are possible, the key here is to minimize the number of operations, and specifically to avoid frequent shuffling of particle data across 256-bit lanes (see Section~\ref{sec:introduction}). 

The following illustration shows the first pairs of accelerations that we calculate.
We have two vectors of particle data to work on, one in which the order is simply sequential and another where the particles have been shuffled.
Specifically, the interactions that we calculate first are between planet 0 and planet 3, planet 1 and planet 2, and so on.
The double-headed arrows, e.g. $0 \leftrightarrow 3$, indicate that we use Newton's third law to calculate the accelerations for both particle~0 and~3 using one square root operation.
The matrix on the right shows in black all 16 accelerations that we calculate in this step.

\begin{center}
\begin{tikzpicture}[scale=0.65]
\tikzstyle{every node}=[font=\fontsize{8}{8}\selectfont]

\def\interactionmatrix{
    {0,2,1,1,3,4,4,3},
    {2,0,1,1,3,3,4,4},
    {1,1,0,2,4,3,3,4},
    {1,1,2,0,4,4,3,3},
    {3,3,4,4,0,2,1,1},
    {4,3,3,4,2,0,1,1},
    {4,4,3,3,1,1,0,2},
    {3,4,4,3,1,1,2,0}% 
}

\begin{scope}[shift={(0,0)}]
    \begin{scope}[shift={(0,2)}]
        \draw[step=10mm, black] (0,0) grid (8,1);
        \foreach \i in {0,...,7} {
            \node at (0.5+\i,0.5) {\i};
        }
    \end{scope}
    \filldraw[fill=black!10!white] (0,0) rectangle (8,1);
    \draw[step=10mm, black] (0,0) grid (8,1);
    \foreach \i [count=\xi] in {3,2,0,1,7,6,4,5} {
        \node at (-0.5+\xi,0.5) {\i};
    }
    \foreach \x in {0,...,7}{   
        \draw[latex-latex] (\x+0.5,1.2) -- (\x+0.5,1.8);
    }   

    \begin{scope}[shift={(9,3)},scale=0.375]
        \foreach \y in {0,...,7} {
           \node[scale=0.8] at (\y+0.5,0.5) {\y};
           \node[scale=0.8] at (-0.5,-\y-0.5) {\y};
        };
        \fill[black]
        \foreach \row [count=\y] in \interactionmatrix {
          \foreach \cell [count=\x] in \row {
            \ifnum\cell=1 %
              (\x-1, -\y+1) rectangle ++(1, -1)
            \fi
            \pgfextra{%
              \global\let\maxx\x
              \global\let\maxy\y
            }%
          }
        };
        \draw[thin] (0, 0) grid[step=1] (\maxx, -\maxy);
    \end{scope}  
\end{scope}
\end{tikzpicture}
\end{center}

\noindent
Next, we shuffle the second of the two vectors in a way that lets us calculate the remaining accelerations in the top left and bottom right quadrants of the matrix.
Note that we do not need to calculate the self-interactions on the diagonal.
Also note that in this step we are calculating the acceleration between each particle pair twice. 
The interaction $0\rightarrow 1$ could be used to calculate $1\rightarrow 0$ but we're calculating it separately. 
Since it takes the same amount of time to calculate eight inverse square roots than it takes to calculate four, this is faster than reusing parts of the calculation which would require another shuffling operation. 
The matrix again shows in black the interactions calculated in this step, and in gray those in the previous step.

\begin{center}
\begin{tikzpicture}[scale=0.65]
\tikzstyle{every node}=[font=\fontsize{8}{8}\selectfont]

\def\interactionmatrix{
    {0,2,1,1,3,4,4,3},
    {2,0,1,1,3,3,4,4},
    {1,1,0,2,4,3,3,4},
    {1,1,2,0,4,4,3,3},
    {3,3,4,4,0,2,1,1},
    {4,3,3,4,2,0,1,1},
    {4,4,3,3,1,1,0,2},
    {3,4,4,3,1,1,2,0}% 
}
%%%%%%%%%%%%%%%%
%%%%%%%%%%%%%%%%
\begin{scope}[shift={(0,-4.5)}]
    \begin{scope}[shift={(0,2)}]
        \draw[step=10mm, black] (0,0) grid (8,1);
        \foreach \i in {0,...,7} {
            \node at (0.5+\i,0.5) {\i};
        }
    \end{scope}
    \filldraw[fill=black!10!white] (0,0) rectangle (8,1);
    \draw[step=10mm, black] (0,0) grid (8,1);
    \node at (0.5,0.5) {1};
    \node at (1.5,0.5) {0};
    \node at (2.5,0.5) {3};
    \node at (3.5,0.5) {2};
    \node at (4.5,0.5) {5};
    \node at (5.5,0.5) {4};
    \node at (6.5,0.5) {7};
    \node at (7.5,0.5) {6};

    \foreach \x in {0,...,7}{   
        \draw[latex-] (\x+0.5,1.2) -- (\x+0.5,1.8);
    }   

    \begin{scope}[shift={(9,3)},scale=0.375]
        \foreach \y in {0,...,7} {
           \node[scale=0.8] at (\y+0.5,0.5) {\y};
           \node[scale=0.8] at (-0.5,-\y-0.5) {\y};
        };
        \fill[black]
        \foreach \row [count=\y] in \interactionmatrix {
          \foreach \cell [count=\x] in \row {
            \ifnum\cell=2 %
              (\x-1, -\y+1) rectangle ++(1, -1)
            \fi
            \pgfextra{%
              \global\let\maxx\x
              \global\let\maxy\y
            }%
          }
        };
        \fill[black!60!white]
        \foreach \row [count=\y] in \interactionmatrix {
          \foreach \cell [count=\x] in \row {
            \ifnum\cell=1 %
              (\x-1, -\y+1) rectangle ++(1, -1)
            \fi
            \pgfextra{%
              \global\let\maxx\x
              \global\let\maxy\y
            }%
          }
        };
        \draw[thin] (0, 0) grid[step=1] (\maxx, -\maxy);
    \end{scope}  
\end{scope}
\end{tikzpicture}
\end{center}

\noindent So far, all the shuffling happened within 256-bit lanes. 
However, the shuffling needed for the next step requires us to cross 256-bit lane boundaries.
The following illustrations show this third step and all the accelerations calculated so far.

%%%%%%%%%%%%%%%%%%%%%%%%%%
\begin{center}
\begin{tikzpicture}[scale=0.65]
\tikzstyle{every node}=[font=\fontsize{8}{8}\selectfont]

\def\interactionmatrix{
    {0,2,1,1,3,4,4,3},
    {2,0,1,1,3,3,4,4},
    {1,1,0,2,4,3,3,4},
    {1,1,2,0,4,4,3,3},
    {3,3,4,4,0,2,1,1},
    {4,3,3,4,2,0,1,1},
    {4,4,3,3,1,1,0,2},
    {3,4,4,3,1,1,2,0}% 
}
%%%%%%%%%%%%%%%%
%%%%%%%%%%%%%%%%
\begin{scope}[shift={(0,-9)}]
    \begin{scope}[shift={(0,2)}]
        \draw[step=10mm, black] (0,0) grid (8,1);
        \foreach \i in {0,...,7} {
            \node at (0.5+\i,0.5) {\i};
        }
    \end{scope}
    \filldraw[fill=black!10!white] (0,0) rectangle (8,1);
    \draw[step=10mm, black] (0,0) grid (8,1);
    \node at (0.5,0.5) {4};
    \node at (1.5,0.5) {5};
    \node at (2.5,0.5) {6};
    \node at (3.5,0.5) {7};
    \node at (4.5,0.5) {1};
    \node at (5.5,0.5) {2};
    \node at (6.5,0.5) {3};
    \node at (7.5,0.5) {0};

    \foreach \x in {0,...,7}{   
        \draw[latex-latex] (\x+0.5,1.2) -- (\x+0.5,1.8);
    }   

    \begin{scope}[shift={(9,3)},scale=0.375]
        \foreach \y in {0,...,7} {
           \node[scale=0.8] at (\y+0.5,0.5) {\y};
           \node[scale=0.8] at (-0.5,-\y-0.5) {\y};
        };
        \fill[black]
        \foreach \row [count=\y] in \interactionmatrix {
          \foreach \cell [count=\x] in \row {
            \ifnum\cell=3 %
              (\x-1, -\y+1) rectangle ++(1, -1)
            \fi
            \pgfextra{%
              \global\let\maxx\x
              \global\let\maxy\y
            }%
          }
        };
        \fill[black!60!white]
        \foreach \row [count=\y] in \interactionmatrix {
          \foreach \cell [count=\x] in \row {
            \ifnum\cell > 0  %
                \ifnum\cell < 3  %
                    (\x-1, -\y+1) rectangle ++(1, -1)
                \fi
            \fi
            \pgfextra{%
              \global\let\maxx\x
              \global\let\maxy\y
            }%
          }
        };
        \draw[thin] (0, 0) grid[step=1] (\maxx, -\maxy);
    \end{scope}  
\end{scope}

\end{tikzpicture}
\end{center}

\noindent 
We need one more shuffling of particle data - this time again within 256-bit lanes - to complete the calculation of all acceleration terms.

%%%%%%%%%%%%%%%%%%%%%%%%%%
\begin{center}
\begin{tikzpicture}[scale=0.65]
\tikzstyle{every node}=[font=\fontsize{8}{8}\selectfont]

\def\interactionmatrix{
    {0,2,1,1,3,4,4,3},
    {2,0,1,1,3,3,4,4},
    {1,1,0,2,4,3,3,4},
    {1,1,2,0,4,4,3,3},
    {3,3,4,4,0,2,1,1},
    {4,3,3,4,2,0,1,1},
    {4,4,3,3,1,1,0,2},
    {3,4,4,3,1,1,2,0}% 
}
%%%%%%%%%%%%%%%%
%%%%%%%%%%%%%%%%
\begin{scope}[shift={(0,-13.5)}]
    \begin{scope}[shift={(0,2)}]
        \draw[step=10mm, black] (0,0) grid (8,1);
        \foreach \i in {0,...,7} {
            \node at (0.5+\i,0.5) {\i};
        }
    \end{scope}
    \filldraw[fill=black!10!white] (0,0) rectangle (8,1);
    \draw[step=10mm, black] (0,0) grid (8,1);
    \node at (0.5,0.5) {5};
    \node at (1.5,0.5) {6};
    \node at (2.5,0.5) {7};
    \node at (3.5,0.5) {4};
    \node at (4.5,0.5) {2};
    \node at (5.5,0.5) {3};
    \node at (6.5,0.5) {0};
    \node at (7.5,0.5) {1};

    \foreach \x in {0,...,7}{   
        \draw[latex-latex] (\x+0.5,1.2) -- (\x+0.5,1.8);
    }   

    \begin{scope}[shift={(9,3)},scale=0.375]
        \foreach \y in {0,...,7} {
           \node[scale=0.8] at (\y+0.5,0.5) {\y};
           \node[scale=0.8] at (-0.5,-\y-0.5) {\y};
        };
        \fill[black]
        \foreach \row [count=\y] in \interactionmatrix {
          \foreach \cell [count=\x] in \row {
            \ifnum\cell=4 %
              (\x-1, -\y+1) rectangle ++(1, -1)
            \fi
            \pgfextra{%
              \global\let\maxx\x
              \global\let\maxy\y
            }%
          }
        };
        \fill[black!60!white]
        \foreach \row [count=\y] in \interactionmatrix {
          \foreach \cell [count=\x] in \row {
            \ifnum\cell > 0  %
                \ifnum\cell < 4  %
                    (\x-1, -\y+1) rectangle ++(1, -1)
                \fi
            \fi
            \pgfextra{%
              \global\let\maxx\x
              \global\let\maxy\y
            }%
          }
        };
        \draw[thin] (0, 0) grid[step=1] (\maxx, -\maxy);
    \end{scope}  
\end{scope}
  
\end{tikzpicture}
\end{center}

\noindent
Now all the time-consuming inverse square-root calculations have been completed. 
However, we need one more shuffling operation (across 256-bit lane boundaries) to combine the accelerations calculated in the first two steps with those in the last two steps. 

In addition to the accelerations in Eq.~\ref{eq:acc}, we also implement an additional acceleration term coming from a $1/r^2$ potential centered on the star. 
This can be used to mimic the effects of general relativistic precession \citep{Nobili1986}.
Because we are working in democratic heliocentric coordinates, the acceleration felt by the planets is easy to calculate in a SIMD fashion without any lane crossings.

However, we also need to include the back reaction onto the star originating from the $1/r^2$ potential. 
For this part, we need to scale the accelerations from all planets, add them, and then subtract the result from the direct contributions. 
This part therefore requires a lane crossing in the same way as the jump step does (see next section).

\subsection{Jump Step}
The so-called jump step in democratic heliocentric coordinates modifies the positions of all planets. 
Specifically, the equations of motions from $H_J$ require us to calculate the sum of the momenta of all planets (but not the star), scale it by some constant factor, and then add it to all planets:
\begin{equation}
   \Delta \mathbf Q_i = \frac{dt}{M_0} \sum_{j=1}^N \mathbf P_j,
\end{equation}
where $M_0$ is the mass of the star.
Note that each planet's change in position depends on the sum of the momenta of all planets. 
By carefully arranging shuffle and add operations, we can implement the jump step with only four additions and three shuffles per coordinate, of which only one shuffle involves a lane crossing across 256-bit boundaries.

Let us assume that some vector $p$ initially contains one Cartesian component of the momenta of all eight planets, $m_0\cdot v_0, m_1\cdot v_1, \ldots, m_7\cdot v_7$. 
We first swap neighbouring pairs of the vector and then add the original vector to the shuffled vector.
The process is repeated two more times, with the second step crossing 128-bit boundaries, and with the last shuffle crossing 256-bit boundaries.
The algorithm is illustrated below: 

\begin{center}
\begin{tikzpicture}[scale=0.75]
    \tikzstyle{every node}=[font=\fontsize{7}{7}\selectfont]
\begin{scope}[shift={(0,2)}]
    \draw[step=10mm, black] (0,0) grid (8,1);
    \node at (0.5,0.5) {0};
    \node at (1.5,0.5) {1};
    \node at (2.5,0.5) {2};
    \node at (3.5,0.5) {3};
    \node at (4.5,0.5) {4};
    \node at (5.5,0.5) {5};
    \node at (6.5,0.5) {6};
    \node at (7.5,0.5) {7};
\end{scope}

\begin{scope}
    \draw[step=10mm, black] (0,0) grid (8,1);
    \node at (0.5,0.5) {1};
    \node at (1.5,0.5) {0};
    \node at (2.5,0.5) {3};
    \node at (3.5,0.5) {2};
    \node at (4.5,0.5) {5};
    \node at (5.5,0.5) {4};
    \node at (6.5,0.5) {7};
    \node at (7.5,0.5) {6};
\end{scope}

\begin{scope}
  \foreach \x in {0,...,3}{   
        \draw[<-] (2*\x+0.7,1.2) -- (2*\x+1.3,1.8);
        \draw[<-] (2*\x+1.3,1.2) -- (2*\x+0.7,1.8);
    }
   \foreach \x in {0,...,7}{ 
        \draw[color=black,fill=white] (\x+0.5,1.5) circle (0.2) node {+};
  }
\end{scope}

\end{tikzpicture}
\begin{tikzpicture}[scale=0.75]
    \tikzstyle{every node}=[font=\fontsize{7}{7}\selectfont]

%%%%%%%%%%%%%%
\draw[<-] (4,-0.8) -- (4,-0.2);
%%%%%%%%%%%%%%

\begin{scope}[shift={(0,-2)}]
    \draw[step=10mm, black] (0,0) grid (8,1);
    \foreach \x in {0,...,1}{   
        \node at (\x+0.5,0.5) {0+1};
        \node at (\x+2.5,0.5) {2+3};
        \node at (\x+4.5,0.5) {4+5};
        \node at (\x+6.5,0.5) {6+7};
    }
\end{scope}

\begin{scope}[shift={(0,-4)}]
    \draw[step=10mm, black] (0,0) grid (8,1);
    \foreach \x in {0,...,1}{   
        \node at (\x+0.5,0.5) {2+3};
        \node at (\x+2.5,0.5) {0+1};
        \node at (\x+4.5,0.5) {6+7};
        \node at (\x+6.5,0.5) {4+5};
    }
\end{scope}

\begin{scope}[shift={(0,-3.5)}]
  \foreach \x in {0,...,1}{   
        \draw[<-] (4*\x+3.3,0.7) -- (4*\x+0.7,1.3);
        \draw[<-] (4*\x+2.3,0.7) -- (4*\x+1.7,1.3);
        \draw[<-] (4*\x+1.7,0.7) -- (4*\x+2.3,1.3);
        \draw[<-] (4*\x+0.7,0.7) -- (4*\x+3.3,1.3);
    }
  \foreach \x in {0,...,7}{ 
        \draw[color=black,fill=white] (\x+0.5,1) circle (0.2) node {+};
  }
\end{scope}

\end{tikzpicture}
\begin{tikzpicture}[scale=0.75]
    \tikzstyle{every node}=[font=\fontsize{7}{7}\selectfont]
%%%%%%%%%%%%%%
\draw[<-] (4,-4.8) -- (4,-4.2);
%%%%%%%%%%%%%%

\begin{scope}[shift={(0,-6)}]
    \draw[step=10mm, black] (0,0) grid (8,1);
    \foreach \x in {0,...,3}{
        \node[align=center] at (\x+0.5,0.5) {0+1\\[-1pt]+\\[-1pt]2+3};
        \node[align=center] at (\x+4.5,0.5) {4+5\\[-1pt]+\\[-1pt]6+7};
    }
\end{scope}

\begin{scope}[shift={(0,-8)}]
    \draw[step=10mm, black] (0,0) grid (8,1);
    \foreach \x in {0,...,3}{
        \node[align=center] at (\x+0.5,0.5) {4+5\\[-1pt]+\\[-1pt]6+7};
        \node[align=center] at (\x+4.5,0.5) {0+1\\[-1pt]+\\[-1pt]2+3};
    }
\end{scope}

\begin{scope}[shift={(0,-7.5)}]
  \foreach \x in {0,...,3}{   
        \draw[<-] (\x+4.3,0.7) -- (\x+0.7,1.3);
        \draw[<-] (\x+0.7,0.7) -- (\x+4.3,1.3);
        
        %\draw[<-] (2*\x+1.3,0.7) -- (2*\x+0.7,1.3);
    }
  \foreach \x in {0,...,7}{ 
        \draw[color=black,fill=white] (\x+0.5,1) circle (0.2) node {+};
  }
\end{scope}

\end{tikzpicture}
\end{center}
In the end, we have the element-wise sum of $p$ in each vector element.
Using Intel's intrinsic functions\footnote{For a list of available intrinsic functions see e.g. \url{https://www.intel.com/content/www/us/en/docs/intrinsics-guide/index.html}.}, this algorithm can be written compactly as
\begin{verbatim}
p = _mm512_add_pd(_mm512_shuffle_pd(
          p, p, 0b01010101),p);
p = _mm512_add_pd(_mm512_permutex_pd(
          p, _MM_PERM_ABCD), p);
p = _mm512_add_pd(_mm512_shuffle_f64x2(
          p,p, _MM_SHUFFLE(1,0,3,2)), p);
\end{verbatim}
During the jump step, the star does not move (there is no back reaction). 

\subsection{Other implementation details}
Throughout WHFast512, we make extensive use of fused multiply-add (FMA) instructions.
As the name suggests, one FMA instruction performs one multiplication and one addition in a single step and with a single rounding at the end.
This not only leads to an increase in performance but also in accuracy.

WHFast512 works with democratic heliocentric coordinates but not with Jacobi coordinates. 
This has the advantage that we do not need to convert back and forth to inertial coordinates during each timestep.
With Jacobi coordinates, coordinate transformations would be required twice per timestep (we need the relative distance between particles in the interaction step which is easier to do in inertial coordinates, but Jacobi coordinates are required for the Kepler step).
Jacobi coordinate transformations are difficult to parallelize in a SIMD fashion because every particle's coordinates depend on the coordinates of all interior particles.
For systems with 8 planets this leads to a significant bottleneck\footnote{Jacobi coordinates would be more feasible for systems with 2 or 4 planets.}. 

\begin{figure*}
    \centering
    \includegraphics[width=\textwidth]{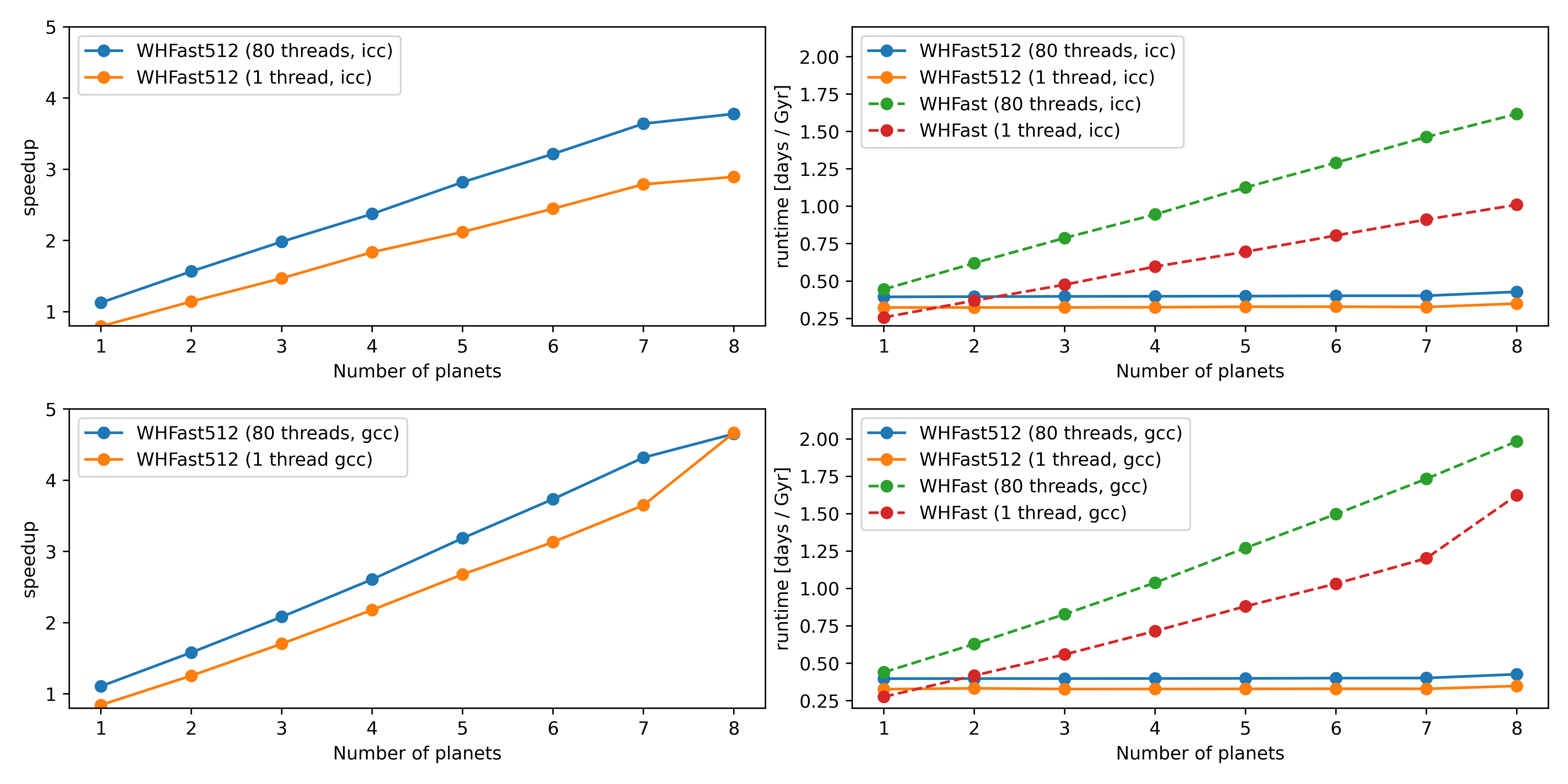}
    \caption{This plot shows the speed up and the runtime as a function of the number of planets. For 8 planets, WHFast512 achieves a speed-up of up to 4.7x.
    A single integration of the Solar System for 1~Gyr takes just 8~hours using a 5-day timestep.
    Note that multi-threading is only used to run multiple independent simulations in parallel, not to speed up one single simulation.
    }
    \label{fig:speedup-parallel}
\end{figure*}

Without loss of generality, we make several further assumptions for WHFast512 to avoid unnecessary operations. 
\begin{itemize}
    \item We assume a fixed timestep. If it becomes necessary to change the timestep at some point during an integration, the user needs to manually reset the integrator. 
    \item The timestep needs to be a positive number. To integrate backwards in time the user can flip the sign of the velocities.
    \item We work in units in which the gravitational constant $G$ is equal to~$1$. Any N-body system can always be rescaled to these units.
    \item We assume that star and planet masses do not change during the integration. If they do, the user needs to manually reset the integrator after each change.
    \item We always combine the two drift steps in the Drift-Kick-Drift algorithm. In the original WHFast version, this is referred to as 'safe mode off'. When an output is required, the integrator needs to be 'synchronized' by applying half a drift step. 
    The 'keep unsynchronized' feature of WHFast is supported, allowing users to synchronize the state vectors to generate an output, but then continue the integration from the unsynchronized values.
    This allows for bit-wise reproducibility independent of the number of outputs and avoids the unnecessary accumulation of round-off errors when frequent outputs are required.
\end{itemize}

\section{Performance}
\label{sec:performance}

In this section, we present results from performance tests.
Specifically, we compare the original non-AVX512 version of WHFast to our new version, WHFast512. 
All simulations use a 5~day timestep and the present day Solar System planets as initial conditions\footnote{Actually, the initial conditions don't matter for this test. We only test the speed, not the accuracy and one WHFast512 timestep always takes the same amount of time regardless of the planets' coordinates.}. 
The tests were performed on the Niagara cluster, owned by the University of Toronto and operated by SciNet. 
Each node has two Intel(R) Xeon(R) Gold 6148 2.40~GHz CPUs (with a maximum frequency 3.7~GHz).
There are a total of 40 Skylake cores per node. 
Using hyperthreading, up to 80 threads can be executed in parallel per node.
We run our tests using both the GNU (version 8.3.0) and Intel compilers (version 19.0.4.243). 
The results for the Intel compiler are shown in the top row of Figure~\ref{fig:speedup-parallel}, and those for the GNU compiler in the bottom row.

We start by running one simulation (one thread) per node. 
This significantly underutilizes the node, but it gives us the fastest possible speed for a single simulation.
For 8 planets, using either Intel or GNU compilers, we are able to integrate for 1~Gyr in just 0.35 days (8.4 hours). 
This is a speedup of 4.7x~(GNU)/2.9x~(Intel) compared to the non-AVX512 version of WHFast. 

Running 80 threads on a single node slightly increases the runtime. 
This is expected because hyperthreading will not provide a perfect scaling, the simulations are more likely memory bound, and the CPU is more likely frequency throttled. 
Note that this is the same for both the original and the AVX512 version of WHFast. 
The speedup for 8 planets is 4.7x/3.7x for the GNU and Intel compiler respectively. 
Looking at the runtime, we can see that to integrate 80 realizations of the Solar system to 1~Gyr on a single node, we need 0.43 days (10 hours) using either the GNU or Intel compilers.
The reason the speed-up is slightly lower for the Intel compiler is that this compiler does some amount of SIMD vectorization by itself.

We also run additional tests on the login nodes of the general purpose Béluga Cluster which uses the same Intel(R) Xeon(R) Gold 6148 CPU but with hyperthreading disabled.
On this machine we achieve an even shorter runtime:
integrating the Solar System for 1~Gyr takes 0.27~days (6.6 hours), allowing us to integrate to 5~Gyr in less than 1.4~days. 

Appendix \ref{sec:other} compares the performance of WHFast512 to several other freely available N-body codes.

\section{Longterm Integrations}
\label{sec:results}
\begin{figure}
    \centering
    \includegraphics[width=\columnwidth]{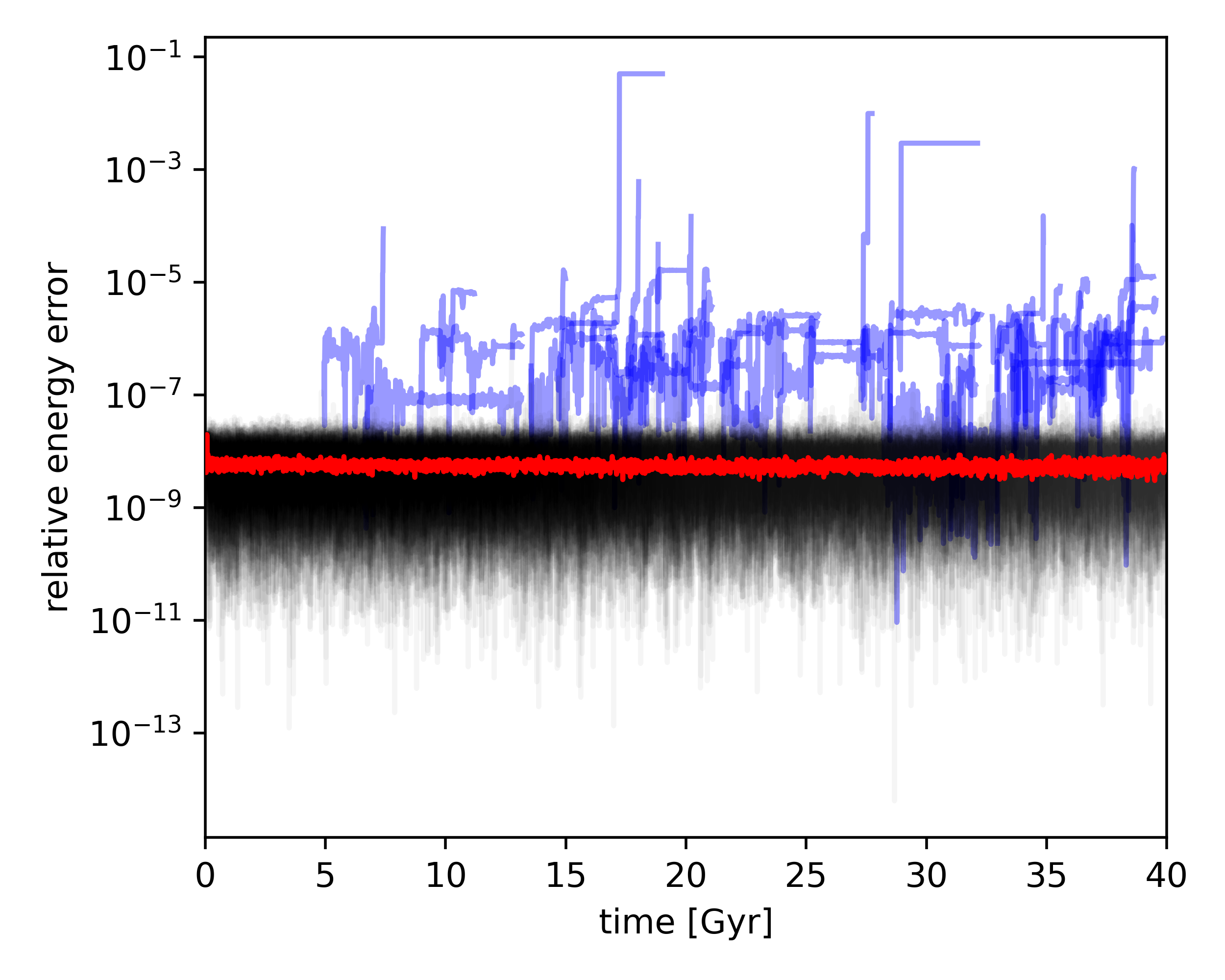}
    \caption{This plot shows the relative energy error in a set of 80 long-term integration of the Solar System with a 5~day timestep and GR corrections. The red curve is the mean of the 80 runs. The blue lines correspond to simulations where Mercury's eccentricity exceeds~$0.65$.
    }
    \label{fig:longterm}
\end{figure}

\begin{figure}
    \centering
    \includegraphics[width=\columnwidth]{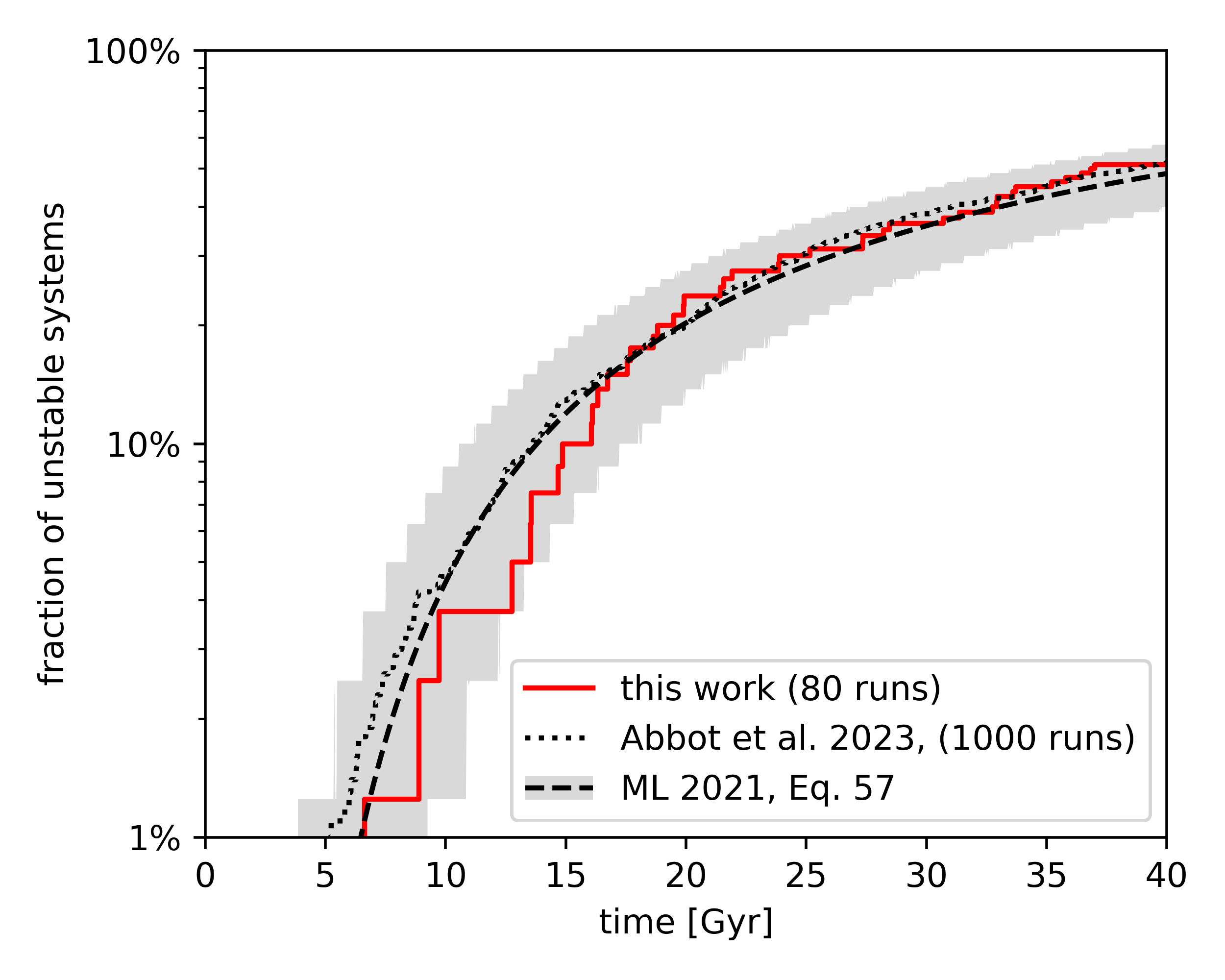}
    \caption{This plot shows the fraction of unstable systems as a function of time. The red curve shows the data from our test runs. Also shown are the N-body results from \cite{Abbot2023b} and the diffusion model estimate of \cite{Mogavero2021} with a $2\sigma$ confidence interval for an ensemble size of 80 integrations.
    }
    \label{fig:unstable}
\end{figure}

We present results from 80 long-term simulations of the Solar System with all 8 planets, general relativistic corrections, and a 5~day timestep.
We start with initial conditions representing the present day Solar System.
The initial conditions of the 80 integrations are identical except that we perturb the $x$~coordinate of Mercury by one meter in a random direction.
Because the Solar System is chaotic, the trajectories of the planets will diverge.

In Figure~\ref{fig:longterm} we plot the relative energy error for our ensemble of simulations. 
Each individual simulation's error is plotted as a gray line while Mercury's eccentricity stays below~$0.65$.
The median is shown in red. 
The error is plotted as a blue line when Mercury's eccentricity exceeds~$0.65$.
One can see that the median relative energy error remains below $10^{-8}$ for the entire 40~Gyr integration. 
The only simulations for which the energy error increases are those where Mercury's eccentricity is high and planets have close encounters. 
Note that there is no long-term trend in the median energy error (red curve), showing that, at least at the level of~$10^{-8}$, the integrator is unbiased over 40~Gyrs. 

The fraction of unstable systems as a function of time in our ensemble runs is shown as a red line in Figure~\ref{fig:unstable}.
\cite{Abbot2023b} ran 1000~simulations using the non-AVX512 version of WHFast.
For comparison, we plot these results as a dotted line.
We also plot the diffusion model by \cite{Mogavero2021} as a dashed line. 
The gray shaded area represents the $2\sigma$ confidence interval for the diffusion model assuming an ensemble size of 80. 

As a criterion for whether a simulation has gone unstable or not we test whether the eccentricity of Mercury exceeds~0.65. 
At this point, the orbits of Mercury and Venus are almost crossing and a violent outcome is pretty much guaranteed. 
We tried other criteria ($e>0.55$, $e>0.6$, relative energy error larger than~$10^{-7}$) all of which give very similar results.

As one can see in Figure~\ref{fig:unstable}, our results are in good agreement with both the diffusion model of \cite{Mogavero2021} and the N-body results of \cite{Abbot2023b}.
This shows that our optimized integrator is well suited to study the long-term evolution of planetary systems such as the Solar System. 

\cite{Wisdom2015} shows that for fewer than~${\sim} 17$ timesteps per pericenter timescale $T_f$, the overlap of timestep resonances introduces energy errors larger than machine precision \citep[see also][]{Rauch1999,Hernandez2022, Abbot2023}. 
Nevertheless, quantifying the impact of such small numerical chaos on the statistics of dynamical instabilities in the Solar System is a challenging theoretical problem. 
Numerical convergence tests by Brown et~al.~(in prep) show that the rate of dynamical instabilities over 5 Gyr in their Solar System integrations remain consistent for timesteps significantly larger than this criterion. 
However, this paper is meant to simply provide a demonstration of our new code, rather than a detailed convergence analysis. 
The fact that our results agree well with the diffusion model, and particularly with the simulations of \cite{Abbot2023b}, which use a 3~day timestep, gives us confidence that our results are physical and not a numerical artifact.

\section{Conclusions}
\label{sec:conclusions}
In this paper, we introduced WHFast512.
To our knowledge, WHFast512 is the fastest N-body integrator for systems with a small number of planets ($N=8$). 
We can integrate the Solar System for 5~billion years, the expected time the Sun has left on the main sequence, in just 1.4~days.

The speedup that we achieve is significant, up to~4.7x. 
It will allow N-body simulations to run almost 5 times longer for the same amount of computing resources. 
Alternatively, one can run 5 times as many simulations for the same integration time and computing resources. 
Lastly, one can run the same number of simulations but at a cost of only 1/5 of the computing resources.
As scientists become more aware of and concerned about the environmental impact of large-scale numerical simulations, the last option seems particularly appealing.

In this paper, we focused on developing an integrator for systems with 8~planets. 
Aside from the Solar System, there are currently few known planetary systems with that many planets. 
It would therefore make sense to further improve the performance of WHFast512 in future work specifically for systems with 2 or 4 planets by sharing one 512-bit vector among 4 or 2 simulations respectively.
In fact, because the interaction matrix is more sparse and there are no longer 256-bit lane crossings required in this case, the speedup should be even greater.
Jacobi coordinates could also be considered for systems with a smaller number of planets (especially 2).
Furthermore, note that some AVX512 instructions can lead to CPU frequency throttling \citep{Lemire2018}. Although we don't observe any significant effect in WHFast512, it would be interesting to compare the performance of 512-bit instructions to twice the amount of 256-bit instructions, especially when integrating 2 or 4 planet systems.

WHFast512 is freely available within the REBOUND package at \url{https://github.com/hannorein/rebound} which provides both a C and Python interface.
To use WHFast512 the user needs to have a CPU and a compiler which support AVX512 instructions.
Note that to allow for easy post-processing of simulation data, one can read SimulationArchive files \citep{ReinTamayo2017} of simulations run with WHFast512 even if AVX512 instructions are not available.
In that case, if a synchronization is required, it is performed with WHFast.

\section*{Acknowledgements}
We thank Dorian Abbot and collaborators for sharing a draft of their paper and the data shown in Figure~\ref{fig:unstable} with us.
We thank Matthew Holman for helpful feedback on an earlier draft of this paper.
We also thank an anonymous referee for helpful comments which allowed us to improve the manuscript.
This research has been supported by the Natural Sciences and Engineering Research Council (NSERC) Discovery Grant RGPIN-2020-04513.
This research was enabled in part by support provided by the Digital Research Alliance of Canada (formerly Compute Canada; \href{https://alliancecan.ca/en}{alliancecan.ca}).
Computations were performed on the Niagara supercomputer \citep{SciNet2010, Ponce2019} at the SciNet HPC Consortium (\url{https://www.scinethpc.ca}). 
SciNet is funded by the following: the Canada Foundation for Innovation; the Government of Ontario; Ontario Research Fund -- Research Excellence; and the University of Toronto.

\appendix

\section{Optimal number of iterations for the Kepler solver}
\label{sec:iterations}

To arrive at the specific combination of Halley's and Newton's method and the number of terms in the Taylor series expansion described in Section~\ref{sec:kepler}, we ran a systematic search over all possible combinations.
As an accuracy requirement we chose that at Mercury's current semi-major axis, the relative semi-major axis error of Mercury stays below $10^{-13}$ when integrated for 10 years with a 5-day timestep and eccentricities up to 0.7. 

Starting with 700 possible combinations, only 85 pass this requirement.
All of them require 6 terms in the Taylor series expansion of the Stumpff functions in the last iteration.
If we only use Newton steps, we need at least six iterations to converge. 
If we start with 1 Halley step, we still need another 4 Newton steps.
If we start with 2 Halley steps, we get away with only 2 more Newton steps. 
The lowest number of terms in the series expansion for the Halley steps that lead to converged results is 4.
Testing the performance of all combinations, this last combination (2 Halley steps, 2 Newton steps, 4 and 6 terms respectively) turns out to be the most efficient and we therefore chose it as the default in WHFast512.
Of course, other requirements - e.g. a different timestep or eccentricity - will lead to a different optimal combination.

\section{Performance compared to other N-body codes}

\begin{table*}[h!]
\centering
\begin{tabular}{l | l l l } 
 \hline
 Package/Integrator & Reference & Wall-time required for 5~Gyr & speed compared to WHFast512  \\ [0.5ex] 
 \hline\hline
 REBOUND/WHFast512 & (this work)        & 1.40 days & - \\ 
 orbitN     & \cite{Zeebe2023}          & 6.48 days & 4.6x slower\\ 
 REBOUND/WHFast & \cite{ReinTamayo2015} & 6.82 days & 4.8x slower\\
 HNBody & \cite{RauchHamilton2002} & 7.17 days & 5.1x slower \\
 swiftest/whm &  \cite{Minton2021} & 13.3 days & 9.5x slower \\
 MERCURY/mvs    & \cite{Chambers1997}   & 46.50 days & 33x slower\\ [1ex] 
 \hline
\end{tabular}
\caption{Speed comparison for different freely available N-body codes. The test problem consists of the Solar System with 8 planets, a 5~day timestep, and general relativistic corrections.}
\label{tab:other}
\end{table*}

\label{sec:other}
We compare WHFast512 primarily to WHFast because both integrators are part of the same package and we think this is the fairest comparison.
In this appendix we report results of experiments comparing WHFast512 to other freely available N-body integrators. 
All of them are open source, except HNBody which is only available as a binary.
Although we took care in making these comparisons as fair as possible and closely follow the developers' instructions, there might be optimizations (settings, compiler flags, etc) in these codes that we did not make use of. 
In addition some of the codes are optimized for other use cases, for example integrations with a large number of test particles or close encounters, rather than integrating the Solar System planets.

Table \ref{tab:other} lists the runtime for a 5~Gyr integration of the eight Solar System planets using a 5~day timestep for different N-body codes. 
All runs were performed on the same Intel(R) Xeon(R) Gold 6148 CPU.

\bibliography{full}

\begin{thebibliography}{26}%
\makeatletter
\providecommand \@ifxundefined [1]{%
 \@ifx{#1\undefined}
}%
\providecommand \@ifnum [1]{%
 \ifnum #1\expandafter \@firstoftwo
 \else \expandafter \@secondoftwo
 \fi
}%
\providecommand \@ifx [1]{%
 \ifx #1\expandafter \@firstoftwo
 \else \expandafter \@secondoftwo
 \fi
}%
\providecommand \natexlab [1]{#1}%
\providecommand \enquote  [1]{``#1''}%
\providecommand \bibnamefont  [1]{#1}%
\providecommand \bibfnamefont [1]{#1}%
\providecommand \citenamefont [1]{#1}%
\providecommand \href@noop [0]{\@secondoftwo}%
\providecommand \href [0]{\begingroup \@sanitize@url \@href}%
\providecommand \@href[1]{\@@startlink{#1}\@@href}%
\providecommand \@@href[1]{\endgroup#1\@@endlink}%
\providecommand \@sanitize@url [0]{\catcode `\\12\catcode `\$12\catcode
  `\&12\catcode `\#12\catcode `\^12\catcode `\_12\catcode `\%12\relax}%
\providecommand \@@startlink[1]{}%
\providecommand \@@endlink[0]{}%
\providecommand \url  [0]{\begingroup\@sanitize@url \@url }%
\providecommand \@url [1]{\endgroup\@href {#1}{\urlprefix }}%
\providecommand \urlprefix  [0]{URL }%
\providecommand \Eprint [0]{\href }%
\providecommand \doibase [0]{http://dx.doi.org/}%
\providecommand \selectlanguage [0]{\@gobble}%
\providecommand \bibinfo  [0]{\@secondoftwo}%
\providecommand \bibfield  [0]{\@secondoftwo}%
\providecommand \translation [1]{[#1]}%
\providecommand \BibitemOpen [0]{}%
\providecommand \bibitemStop [0]{}%
\providecommand \bibitemNoStop [0]{.\EOS\space}%
\providecommand \EOS [0]{\spacefactor3000\relax}%
\providecommand \BibitemShut  [1]{\csname bibitem#1\endcsname}%
\let\auto@bib@innerbib\@empty
%</preamble>
\bibitem [{\citenamefont {Laskar}(2013)}]{Laskar2013}%
  \BibitemOpen
  \bibfield  {author} {\bibinfo {author} {\bibfnamefont {J.}~\bibnamefont
  {Laskar}},\ }in\ \href@noop {} {\emph {\bibinfo {booktitle} {Chaos}}}\
  (\bibinfo  {publisher} {Springer},\ \bibinfo {year} {2013})\ pp.\ \bibinfo
  {pages} {239--270},\ \Eprint {http://arxiv.org/abs/1209.5996}
  {arXiv:1209.5996 [astro-ph.EP]} \BibitemShut {NoStop}%
\bibitem [{\citenamefont {{Wisdom}}(1981)}]{Wisdom1981}%
  \BibitemOpen
  \bibfield  {author} {\bibinfo {author} {\bibfnamefont {J.~L.}\ \bibnamefont
  {{Wisdom}}},\ }\emph {\bibinfo {title} {{The origin of the Kirkwood gaps: A
  mapping for asteroidal motion near the 3/1 commensurability 1. The resonance
  overlap criterion and the onset of stochastic behavior in the restricted
  three-body problem 2.}}},\ \href@noop {} {Ph.D. thesis},\ \bibinfo  {school}
  {California Institute of Technology, Pasadena.} (\bibinfo {year}
  {1981})\BibitemShut {NoStop}%
\bibitem [{\citenamefont {{Wisdom}}\ and\ \citenamefont
  {{Holman}}(1991)}]{WisdomHolman1991}%
  \BibitemOpen
  \bibfield  {author} {\bibinfo {author} {\bibfnamefont {J.}~\bibnamefont
  {{Wisdom}}}\ and\ \bibinfo {author} {\bibfnamefont {M.}~\bibnamefont
  {{Holman}}},\ }\href {\doibase 10.1086/115978} {\bibfield  {journal}
  {\bibinfo  {journal} {\aj}\ }\textbf {\bibinfo {volume} {102}},\ \bibinfo
  {pages} {1528} (\bibinfo {year} {1991})}\BibitemShut {NoStop}%
\bibitem [{\citenamefont {{Kinoshita}}\ \emph {et~al.}(1991)\citenamefont
  {{Kinoshita}}, \citenamefont {{Yoshida}},\ and\ \citenamefont
  {{Nakai}}}]{Kinoshita1991}%
  \BibitemOpen
  \bibfield  {author} {\bibinfo {author} {\bibfnamefont {H.}~\bibnamefont
  {{Kinoshita}}}, \bibinfo {author} {\bibfnamefont {H.}~\bibnamefont
  {{Yoshida}}}, \ and\ \bibinfo {author} {\bibfnamefont {H.}~\bibnamefont
  {{Nakai}}},\ }\href@noop {} {\bibfield  {journal} {\bibinfo  {journal}
  {Celestial Mechanics and Dynamical Astronomy}\ }\textbf {\bibinfo {volume}
  {50}},\ \bibinfo {pages} {59} (\bibinfo {year} {1991})}\BibitemShut {NoStop}%
\bibitem [{\citenamefont {{Rein}}\ and\ \citenamefont
  {{Tamayo}}(2015)}]{ReinTamayo2015}%
  \BibitemOpen
  \bibfield  {author} {\bibinfo {author} {\bibfnamefont {H.}~\bibnamefont
  {{Rein}}}\ and\ \bibinfo {author} {\bibfnamefont {D.}~\bibnamefont
  {{Tamayo}}},\ }\href@noop {} {\bibfield  {journal} {\bibinfo  {journal}
  {MNRAS}\ }\textbf {\bibinfo {volume} {452}},\ \bibinfo {pages} {376}
  (\bibinfo {year} {2015})}\BibitemShut {NoStop}%
\bibitem [{\citenamefont {{Rein}}\ and\ \citenamefont
  {{Liu}}(2012)}]{ReinLiu2012}%
  \BibitemOpen
  \bibfield  {author} {\bibinfo {author} {\bibfnamefont {H.}~\bibnamefont
  {{Rein}}}\ and\ \bibinfo {author} {\bibfnamefont {S.~F.}\ \bibnamefont
  {{Liu}}},\ }\href {\doibase 10.1051/0004-6361/201118085} {\bibfield
  {journal} {\bibinfo  {journal} {\aap}\ }\textbf {\bibinfo {volume} {537}},\
  \bibinfo {eid} {A128} (\bibinfo {year} {2012})},\ \Eprint
  {http://arxiv.org/abs/1110.4876} {arXiv:1110.4876 [astro-ph.EP]} \BibitemShut
  {NoStop}%
\bibitem [{\citenamefont {{Grimm}}\ \emph {et~al.}(2022)\citenamefont
  {{Grimm}}, \citenamefont {{Stadel}}, \citenamefont {{Brasser}}, \citenamefont
  {{Meier}},\ and\ \citenamefont {{Mordasini}}}]{Grimm2022}%
  \BibitemOpen
  \bibfield  {author} {\bibinfo {author} {\bibfnamefont {S.~L.}\ \bibnamefont
  {{Grimm}}}, \bibinfo {author} {\bibfnamefont {J.~G.}\ \bibnamefont
  {{Stadel}}}, \bibinfo {author} {\bibfnamefont {R.}~\bibnamefont {{Brasser}}},
  \bibinfo {author} {\bibfnamefont {M.~M.~M.}\ \bibnamefont {{Meier}}}, \ and\
  \bibinfo {author} {\bibfnamefont {C.}~\bibnamefont {{Mordasini}}},\ }\href
  {\doibase 10.3847/1538-4357/ac6dd2} {\bibfield  {journal} {\bibinfo
  {journal} {\apj}\ }\textbf {\bibinfo {volume} {932}},\ \bibinfo {eid} {124}
  (\bibinfo {year} {2022})},\ \Eprint {http://arxiv.org/abs/2201.10058}
  {arXiv:2201.10058 [astro-ph.EP]} \BibitemShut {NoStop}%
\bibitem [{\citenamefont {{Saha}}\ \emph {et~al.}(1997)\citenamefont {{Saha}},
  \citenamefont {{Stadel}},\ and\ \citenamefont
  {{Tremaine}}}]{SahaStadelTremaine1997}%
  \BibitemOpen
  \bibfield  {author} {\bibinfo {author} {\bibfnamefont {P.}~\bibnamefont
  {{Saha}}}, \bibinfo {author} {\bibfnamefont {J.}~\bibnamefont {{Stadel}}}, \
  and\ \bibinfo {author} {\bibfnamefont {S.}~\bibnamefont {{Tremaine}}},\
  }\href {\doibase 10.1086/118485} {\bibfield  {journal} {\bibinfo  {journal}
  {\aj}\ }\textbf {\bibinfo {volume} {114}},\ \bibinfo {pages} {409} (\bibinfo
  {year} {1997})},\ \Eprint {http://arxiv.org/abs/arXiv:astro-ph/9605016}
  {arXiv:astro-ph/9605016} \BibitemShut {NoStop}%
\bibitem [{\citenamefont {{Duncan}}\ \emph {et~al.}(1998)\citenamefont
  {{Duncan}}, \citenamefont {{Levison}},\ and\ \citenamefont
  {{Lee}}}]{Duncan1998}%
  \BibitemOpen
  \bibfield  {author} {\bibinfo {author} {\bibfnamefont {M.~J.}\ \bibnamefont
  {{Duncan}}}, \bibinfo {author} {\bibfnamefont {H.~F.}\ \bibnamefont
  {{Levison}}}, \ and\ \bibinfo {author} {\bibfnamefont {M.~H.}\ \bibnamefont
  {{Lee}}},\ }\href {\doibase 10.1086/300541} {\bibfield  {journal} {\bibinfo
  {journal} {\aj}\ }\textbf {\bibinfo {volume} {116}},\ \bibinfo {pages} {2067}
  (\bibinfo {year} {1998})}\BibitemShut {NoStop}%
\bibitem [{\citenamefont {{Hernandez}}\ and\ \citenamefont
  {{Dehnen}}(2017)}]{HernandezDehnen2016}%
  \BibitemOpen
  \bibfield  {author} {\bibinfo {author} {\bibfnamefont {D.~M.}\ \bibnamefont
  {{Hernandez}}}\ and\ \bibinfo {author} {\bibfnamefont {W.}~\bibnamefont
  {{Dehnen}}},\ }\href {\doibase 10.1093/mnras/stx547} {\bibfield  {journal}
  {\bibinfo  {journal} {Monthly Notices of the Royal Astronomical Society}\
  }\textbf {\bibinfo {volume} {468}},\ \bibinfo {pages} {2614} (\bibinfo {year}
  {2017})},\ \Eprint {http://arxiv.org/abs/1612.05329} {arXiv:1612.05329
  [astro-ph.EP]} \BibitemShut {NoStop}%
\bibitem [{\citenamefont {Rein}\ and\ \citenamefont
  {Tamayo}(2019)}]{ReinTamayo2019}%
  \BibitemOpen
  \bibfield  {author} {\bibinfo {author} {\bibfnamefont {H.}~\bibnamefont
  {Rein}}\ and\ \bibinfo {author} {\bibfnamefont {D.}~\bibnamefont {Tamayo}},\
  }\href {\doibase 10.3847/2515-5172/aaff63} {\bibfield  {journal} {\bibinfo
  {journal} {Research Notes of the {AAS}}\ }\textbf {\bibinfo {volume} {3}},\
  \bibinfo {pages} {16} (\bibinfo {year} {2019})}\BibitemShut {NoStop}%
\bibitem [{\citenamefont {{Wisdom}}(2015)}]{Wisdom2015}%
  \BibitemOpen
  \bibfield  {author} {\bibinfo {author} {\bibfnamefont {J.}~\bibnamefont
  {{Wisdom}}},\ }\href {\doibase 10.1088/0004-6256/150/4/127} {\bibfield
  {journal} {\bibinfo  {journal} {\aj}\ }\textbf {\bibinfo {volume} {150}},\
  \bibinfo {eid} {127} (\bibinfo {year} {2015})}\BibitemShut {NoStop}%
\bibitem [{\citenamefont {{Nobili}}\ and\ \citenamefont
  {{Roxburgh}}(1986)}]{Nobili1986}%
  \BibitemOpen
  \bibfield  {author} {\bibinfo {author} {\bibfnamefont {A.}~\bibnamefont
  {{Nobili}}}\ and\ \bibinfo {author} {\bibfnamefont {I.~W.}\ \bibnamefont
  {{Roxburgh}}},\ }in\ \href@noop {} {\emph {\bibinfo {booktitle} {Relativity
  in Celestial Mechanics and Astrometry. High Precision Dynamical Theories and
  Observational Verifications}}},\ \bibinfo {series} {IAU Symposium}, Vol.\
  \bibinfo {volume} {114},\ \bibinfo {editor} {edited by\ \bibinfo {editor}
  {\bibfnamefont {J.}~\bibnamefont {{Kovalevsky}}}\ and\ \bibinfo {editor}
  {\bibfnamefont {V.~A.}\ \bibnamefont {{Brumberg}}}}\ (\bibinfo {year}
  {1986})\ pp.\ \bibinfo {pages} {105--110}\BibitemShut {NoStop}%
\bibitem [{\citenamefont {{Abbot}}\ \emph
  {et~al.}(2023{\natexlab{a}})\citenamefont {{Abbot}}, \citenamefont
  {{Webber}}, \citenamefont {{Hernandez}}, \citenamefont {{Hadden}},\ and\
  \citenamefont {{Weare}}}]{Abbot2023b}%
  \BibitemOpen
  \bibfield  {author} {\bibinfo {author} {\bibfnamefont {D.~S.}\ \bibnamefont
  {{Abbot}}}, \bibinfo {author} {\bibfnamefont {R.~J.}\ \bibnamefont
  {{Webber}}}, \bibinfo {author} {\bibfnamefont {D.~M.}\ \bibnamefont
  {{Hernandez}}}, \bibinfo {author} {\bibfnamefont {S.}~\bibnamefont
  {{Hadden}}}, \ and\ \bibinfo {author} {\bibfnamefont {J.}~\bibnamefont
  {{Weare}}},\ }\href {\doibase 10.48550/arXiv.2306.11870} {\bibfield
  {journal} {\bibinfo  {journal} {arXiv e-prints}\ ,\ \bibinfo {eid}
  {arXiv:2306.11870}} (\bibinfo {year} {2023}{\natexlab{a}})},\ \Eprint
  {http://arxiv.org/abs/2306.11870} {arXiv:2306.11870 [astro-ph.EP]}
  \BibitemShut {NoStop}%
\bibitem [{\citenamefont {{Mogavero}}\ and\ \citenamefont
  {{Laskar}}(2021)}]{Mogavero2021}%
  \BibitemOpen
  \bibfield  {author} {\bibinfo {author} {\bibfnamefont {F.}~\bibnamefont
  {{Mogavero}}}\ and\ \bibinfo {author} {\bibfnamefont {J.}~\bibnamefont
  {{Laskar}}},\ }\href {\doibase 10.1051/0004-6361/202141007} {\bibfield
  {journal} {\bibinfo  {journal} {\aap}\ }\textbf {\bibinfo {volume} {655}},\
  \bibinfo {eid} {A1} (\bibinfo {year} {2021})},\ \Eprint
  {http://arxiv.org/abs/2105.14976} {arXiv:2105.14976 [astro-ph.EP]}
  \BibitemShut {NoStop}%
\bibitem [{\citenamefont {Rauch}\ and\ \citenamefont
  {Holman}(1999)}]{Rauch1999}%
  \BibitemOpen
  \bibfield  {author} {\bibinfo {author} {\bibfnamefont {K.~P.}\ \bibnamefont
  {Rauch}}\ and\ \bibinfo {author} {\bibfnamefont {M.}~\bibnamefont {Holman}},\
  }\href@noop {} {\bibfield  {journal} {\bibinfo  {journal} {The Astronomical
  Journal}\ }\textbf {\bibinfo {volume} {117}},\ \bibinfo {pages} {1087}
  (\bibinfo {year} {1999})}\BibitemShut {NoStop}%
\bibitem [{\citenamefont {{Hernandez}}\ \emph {et~al.}(2022)\citenamefont
  {{Hernandez}}, \citenamefont {{Zeebe}},\ and\ \citenamefont
  {{Hadden}}}]{Hernandez2022}%
  \BibitemOpen
  \bibfield  {author} {\bibinfo {author} {\bibfnamefont {D.~M.}\ \bibnamefont
  {{Hernandez}}}, \bibinfo {author} {\bibfnamefont {R.~E.}\ \bibnamefont
  {{Zeebe}}}, \ and\ \bibinfo {author} {\bibfnamefont {S.}~\bibnamefont
  {{Hadden}}},\ }\href {\doibase 10.1093/mnras/stab3664} {\bibfield  {journal}
  {\bibinfo  {journal} {\mnras}\ }\textbf {\bibinfo {volume} {510}},\ \bibinfo
  {pages} {4302} (\bibinfo {year} {2022})},\ \Eprint
  {http://arxiv.org/abs/2111.08835} {arXiv:2111.08835 [astro-ph.EP]}
  \BibitemShut {NoStop}%
\bibitem [{\citenamefont {{Abbot}}\ \emph
  {et~al.}(2023{\natexlab{b}})\citenamefont {{Abbot}}, \citenamefont
  {{Hernandez}}, \citenamefont {{Hadden}}, \citenamefont {{Webber}},
  \citenamefont {{Afentakis}},\ and\ \citenamefont {{Weare}}}]{Abbot2023}%
  \BibitemOpen
  \bibfield  {author} {\bibinfo {author} {\bibfnamefont {D.~S.}\ \bibnamefont
  {{Abbot}}}, \bibinfo {author} {\bibfnamefont {D.~M.}\ \bibnamefont
  {{Hernandez}}}, \bibinfo {author} {\bibfnamefont {S.}~\bibnamefont
  {{Hadden}}}, \bibinfo {author} {\bibfnamefont {R.~J.}\ \bibnamefont
  {{Webber}}}, \bibinfo {author} {\bibfnamefont {G.~P.}\ \bibnamefont
  {{Afentakis}}}, \ and\ \bibinfo {author} {\bibfnamefont {J.}~\bibnamefont
  {{Weare}}},\ }\href {\doibase 10.3847/1538-4357/acb6ff} {\bibfield  {journal}
  {\bibinfo  {journal} {\apj}\ }\textbf {\bibinfo {volume} {944}},\ \bibinfo
  {eid} {190} (\bibinfo {year} {2023}{\natexlab{b}})},\ \Eprint
  {http://arxiv.org/abs/2212.14844} {arXiv:2212.14844 [astro-ph.EP]}
  \BibitemShut {NoStop}%
\bibitem [{\citenamefont {Lemire}(2018)}]{Lemire2018}%
  \BibitemOpen
  \bibfield  {author} {\bibinfo {author} {\bibfnamefont {D.}~\bibnamefont
  {Lemire}},\ }\href
  {https://web.archive.org/web/20230320071456/https://lemire.me/blog/2018/08/25/avx-512-throttling-heavy-instructions-are-maybe-not-so-dangerous/}
  {\enquote {\bibinfo {title} {Avx-512 throttling: heavy instructions are maybe
  not so dangerous},}\ } (\bibinfo {year} {2018}),\ \bibinfo {note}
  {\url{https://lemire.me/blog/2018/08/25/avx-512-throttling-heavy-instructions-are-maybe-not-so-dangerous/}
  [accessed Jul 19 2023]}\BibitemShut {NoStop}%
\bibitem [{\citenamefont {{Rein}}\ and\ \citenamefont
  {{Tamayo}}(2017)}]{ReinTamayo2017}%
  \BibitemOpen
  \bibfield  {author} {\bibinfo {author} {\bibfnamefont {H.}~\bibnamefont
  {{Rein}}}\ and\ \bibinfo {author} {\bibfnamefont {D.}~\bibnamefont
  {{Tamayo}}},\ }\href@noop {} {\bibfield  {journal} {\bibinfo  {journal}
  {MNRAS}\ }\textbf {\bibinfo {volume} {467}},\ \bibinfo {pages} {2377}
  (\bibinfo {year} {2017})}\BibitemShut {NoStop}%
\bibitem [{\citenamefont {Loken}\ \emph {et~al.}(2010)\citenamefont {Loken},
  \citenamefont {Gruner}, \citenamefont {Groer}, \citenamefont {Peltier},
  \citenamefont {Bunn}, \citenamefont {Craig}, \citenamefont {Henriques},
  \citenamefont {Dempsey}, \citenamefont {Yu}, \citenamefont {Chen},
  \citenamefont {Dursi}, \citenamefont {Chong}, \citenamefont {Northrup},
  \citenamefont {Pinto}, \citenamefont {Knecht},\ and\ \citenamefont
  {Zon}}]{SciNet2010}%
  \BibitemOpen
  \bibfield  {author} {\bibinfo {author} {\bibfnamefont {C.}~\bibnamefont
  {Loken}}, \bibinfo {author} {\bibfnamefont {D.}~\bibnamefont {Gruner}},
  \bibinfo {author} {\bibfnamefont {L.}~\bibnamefont {Groer}}, \bibinfo
  {author} {\bibfnamefont {R.}~\bibnamefont {Peltier}}, \bibinfo {author}
  {\bibfnamefont {N.}~\bibnamefont {Bunn}}, \bibinfo {author} {\bibfnamefont
  {M.}~\bibnamefont {Craig}}, \bibinfo {author} {\bibfnamefont
  {T.}~\bibnamefont {Henriques}}, \bibinfo {author} {\bibfnamefont
  {J.}~\bibnamefont {Dempsey}}, \bibinfo {author} {\bibfnamefont {C.-H.}\
  \bibnamefont {Yu}}, \bibinfo {author} {\bibfnamefont {J.}~\bibnamefont
  {Chen}}, \bibinfo {author} {\bibfnamefont {L.~J.}\ \bibnamefont {Dursi}},
  \bibinfo {author} {\bibfnamefont {J.}~\bibnamefont {Chong}}, \bibinfo
  {author} {\bibfnamefont {S.}~\bibnamefont {Northrup}}, \bibinfo {author}
  {\bibfnamefont {J.}~\bibnamefont {Pinto}}, \bibinfo {author} {\bibfnamefont
  {N.}~\bibnamefont {Knecht}}, \ and\ \bibinfo {author} {\bibfnamefont {R.~V.}\
  \bibnamefont {Zon}},\ }\href {\doibase 10.1088/1742-6596/256/1/012026}
  {\bibfield  {journal} {\bibinfo  {journal} {Journal of Physics: Conference
  Series}\ }\textbf {\bibinfo {volume} {256}},\ \bibinfo {pages} {012026}
  (\bibinfo {year} {2010})}\BibitemShut {NoStop}%
\bibitem [{\citenamefont {Ponce}\ \emph {et~al.}(2019)\citenamefont {Ponce},
  \citenamefont {van Zon}, \citenamefont {Northrup}, \citenamefont {Gruner},
  \citenamefont {Chen}, \citenamefont {Ertinaz}, \citenamefont {Fedoseev},
  \citenamefont {Groer}, \citenamefont {Mao}, \citenamefont {Mundim} \emph
  {et~al.}}]{Ponce2019}%
  \BibitemOpen
  \bibfield  {author} {\bibinfo {author} {\bibfnamefont {M.}~\bibnamefont
  {Ponce}}, \bibinfo {author} {\bibfnamefont {R.}~\bibnamefont {van Zon}},
  \bibinfo {author} {\bibfnamefont {S.}~\bibnamefont {Northrup}}, \bibinfo
  {author} {\bibfnamefont {D.}~\bibnamefont {Gruner}}, \bibinfo {author}
  {\bibfnamefont {J.}~\bibnamefont {Chen}}, \bibinfo {author} {\bibfnamefont
  {F.}~\bibnamefont {Ertinaz}}, \bibinfo {author} {\bibfnamefont
  {A.}~\bibnamefont {Fedoseev}}, \bibinfo {author} {\bibfnamefont
  {L.}~\bibnamefont {Groer}}, \bibinfo {author} {\bibfnamefont
  {F.}~\bibnamefont {Mao}}, \bibinfo {author} {\bibfnamefont {B.~C.}\
  \bibnamefont {Mundim}},  \emph {et~al.},\ }in\ \href@noop {} {\emph {\bibinfo
  {booktitle} {Proceedings of the Practice and Experience in Advanced Research
  Computing on Rise of the Machines (learning)}}}\ (\bibinfo {organization}
  {ACM},\ \bibinfo {year} {2019})\ p.~\bibinfo {pages} {34}\BibitemShut
  {NoStop}%
\bibitem [{\citenamefont {Zeebe}(2023)}]{Zeebe2023}%
  \BibitemOpen
  \bibfield  {author} {\bibinfo {author} {\bibfnamefont {R.~E.}\ \bibnamefont
  {Zeebe}},\ }\href {\doibase 10.3847/1538-3881/acd63b} {\bibfield  {journal}
  {\bibinfo  {journal} {The Astronomical Journal}\ }\textbf {\bibinfo {volume}
  {166}},\ \bibinfo {pages} {1} (\bibinfo {year} {2023})}\BibitemShut {NoStop}%
\bibitem [{\citenamefont {{Rauch}}\ and\ \citenamefont
  {{Hamilton}}(2002)}]{RauchHamilton2002}%
  \BibitemOpen
  \bibfield  {author} {\bibinfo {author} {\bibfnamefont {K.~P.}\ \bibnamefont
  {{Rauch}}}\ and\ \bibinfo {author} {\bibfnamefont {D.~P.}\ \bibnamefont
  {{Hamilton}}},\ }in\ \href@noop {} {\emph {\bibinfo {booktitle} {AAS/Division
  of Dynamical Astronomy Meeting \#33}}},\ \bibinfo {series} {AAS/Division of
  Dynamical Astronomy Meeting}, Vol.~\bibinfo {volume} {33}\ (\bibinfo {year}
  {2002})\ p.\ \bibinfo {pages} {08.02}\BibitemShut {NoStop}%
\bibitem [{\citenamefont {{Minton}}\ \emph {et~al.}(2021)\citenamefont
  {{Minton}}, \citenamefont {{Wishard}}, \citenamefont {{Pouplin}},
  \citenamefont {{Elliott}},\ and\ \citenamefont {{Singh}}}]{Minton2021}%
  \BibitemOpen
  \bibfield  {author} {\bibinfo {author} {\bibfnamefont {D.}~\bibnamefont
  {{Minton}}}, \bibinfo {author} {\bibfnamefont {C.}~\bibnamefont {{Wishard}}},
  \bibinfo {author} {\bibfnamefont {J.}~\bibnamefont {{Pouplin}}}, \bibinfo
  {author} {\bibfnamefont {J.}~\bibnamefont {{Elliott}}}, \ and\ \bibinfo
  {author} {\bibfnamefont {D.}~\bibnamefont {{Singh}}},\ }in\ \href@noop {}
  {\emph {\bibinfo {booktitle} {AAS/Division for Planetary Sciences Meeting
  Abstracts}}},\ \bibinfo {series} {AAS/Division for Planetary Sciences Meeting
  Abstracts}, Vol.~\bibinfo {volume} {53}\ (\bibinfo {year} {2021})\ p.\
  \bibinfo {pages} {411.03}\BibitemShut {NoStop}%
\bibitem [{\citenamefont {{Chambers}}\ and\ \citenamefont
  {{Migliorini}}(1997)}]{Chambers1997}%
  \BibitemOpen
  \bibfield  {author} {\bibinfo {author} {\bibfnamefont {J.~E.}\ \bibnamefont
  {{Chambers}}}\ and\ \bibinfo {author} {\bibfnamefont {F.}~\bibnamefont
  {{Migliorini}}},\ }in\ \href@noop {} {\emph {\bibinfo {booktitle}
  {AAS/Division for Planetary Sciences Meeting Abstracts \#29}}},\ \bibinfo
  {series} {Bulletin of the American Astronomical Society}, Vol.~\bibinfo
  {volume} {29}\ (\bibinfo {year} {1997})\ p.\ \bibinfo {pages}
  {1024}\BibitemShut {NoStop}%
\end{thebibliography}%

\onecolumngrid \footnotesize

\end{document}